\documentclass[aps,prb,reprint]{revtex4-2}
\usepackage{graphicx}
\usepackage{graphics}
\usepackage{amsfonts}
\usepackage{amsmath}
\usepackage{amssymb}
\usepackage{bm}
\usepackage{color}

\begin{document}

\title{Classical diamagnetism: How adiabatic change of field makes it work}

\author{Ya.\ B. Bazaliy}
  \email{bazaliy@mailbox.sc.edu}
  \affiliation{Department of Physics and Astronomy, University of South Carolina, Columbia, SC 29208, USA}

\author{Ammar Kasem}
  \email{akasem@email.sc.edu}
  \affiliation{Department of Physics and Astronomy, University of South Carolina, Columbia, SC 29208, USA}

\date{\today}

\begin{abstract}
Many textbooks discuss the diamagnetic response of a ``classical atom'' to a small and slowly (adiabatically) applied magnetic field. Here we show that the small-field assumption is not essential, whereas the slow rate of field change is crucial: it preserves the circular nature of the orbit regardless of the applied field magnitude. The latter property follows from the existence of an adiabatic invariant of electron's motion. Orbit circularity allows us to solve the problem for arbitrarily large fields. Resulting diamagnetic response exhibits symmetries that unexpectedly persist beyond the small-field limit, and provides a nice illustration of the range of applicability of Larmor's theorem.

\end{abstract}

\maketitle

\section{Introduction}
The model of a classical diamagnetic atom is a part of the standard curriculum in electrodynamics \cite{textbook_griffiths, textbook_purcell}. While the Bohr-Van Leeuwen theorem \cite{feynman_lectures, vanvleck_book_1932, textbook_aharoni, textbook_blundell} and the radiative loss of the orbiting electron energy (Ref.~\onlinecite{textbook_griffiths}, Sec. 11.2.2)  independently show that classical description cannot account for natural diamagnetism, this model still helps students develop an important intuition about the orbital response to magnetic fields, and therefore retains considerable pedagogical value.

\begin{figure}[t]
\center
\includegraphics[width = 0.25\textwidth]{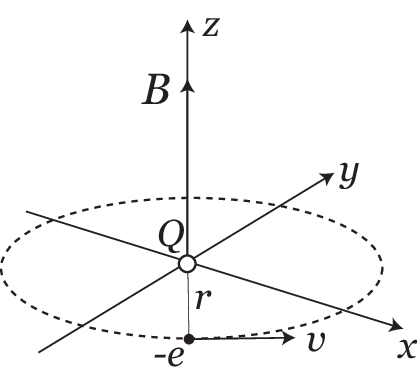}
\caption{Classical model of a diamagnetic atom.}
\label{fig:atom}
\end{figure}

In the classical model electron's motion is assumed to be governed by the non-relativistic Newton's equations. Atomic electron initially moves along a circular orbit in the $(x,y)$ plane with centripetal acceleration provided by the Coulomb attraction to the nucleus (Fig.~\ref{fig:atom}). Circular motion produces a magnetic moment ${\vec m}_0 = m_0 \hat z$. Then, a uniform external magnetic field $\vec B(t) = B(t) \hat z$, directed perpendicular to the plane of the orbit, is slowly turned on. Magnetic field variation induces a circular electric field $\vec E_{ind}(t)$. The two fields, acting together, modify the electron's trajectory and change its magnetic moment, resulting in the diamagnetic response of the atom.

Textbooks discuss the field-induced magnetic moment $m(B)$ under two assumptions. The first requires a slow, ``adiabatic'' change of the field with $\dot B(t) \to 0$. The second demands that throughout the process $B(t)$ itself remains small in the sense that Lorentz force acting on the electron is small compared with the Coulomb force. They obtain the relation (in CGS)
\begin{equation}\label{eq:m_small_B}
m(B) = m_0 - \frac{e^2 r_0^2}{4 c m} B \ ,
\end{equation}
where $(-e) < 0$ is the electron's charge, $m$ is the electron's mass, $c$ is the speed of light, and $r_0$ is the initial radius of the orbit \cite{textbook_griffiths}. Formula (\ref{eq:m_small_B}) predicts a magnetic moment change that is independent of the history of magnetic field variation---as long as $B(t)$ changes adiabatically. The assumption of smallness of $B$ is essential for the derivation: it allows one to neglect the change of the radius of electron's orbit and set $r(B) = r_0$. That property is postulated by some authors \cite{feynman_lectures}, justified in clever ways by others \cite{textbook_purcell,odell_ajp1986,boyer_ajp2019}, or left for the students to prove \cite{textbook_griffiths}.

In the latter case, an avid student finds that the constancy of the radius is not an exact statement but rather an approximation. All one can show is that for $B \to 0$ the change of magnetic moment is linear in applied field $\Delta m = m(B) - m_0 \sim B$, while the change of the radius $\Delta r = r(B) - r_0$ is of higher order in $B$. Having this property is enough to justify the linear expression (\ref{eq:m_small_B}). This seems to put the matter to rest. However, an even closer inspection shows that yet another assumption is implicitly made in Refs. \onlinecite{textbook_purcell,feynman_lectures,odell_ajp1986,boyer_ajp2019,textbook_griffiths} without justification. Namely, one may still question the statement that the form of the orbit remains circular as magnetic field is turned on. Indeed, even in the absence of $\vec B$ and $\vec E_{ind}$ a classical electron moving in the Coulomb field of the nucleus may follow infinitely many elliptic orbits. Thus one can legitimately worry that even a small external perturbation, especially when acting during many revolution periods, may considerably modify the circular nature of the original orbit. If that happens, $r$ would vary during the revolution period, and $r(B)$ will not be well-defined.

Here we explain why in the limit of infinitely slow application of magnetic field, $\dot B \to 0$, the orbit remains circular, and for a small but finite rate of field change the trajectory only slightly deviates from a circle. We also demonstrate the converse: even a small but fast change of $B$ produces orbits for which the fluctuation of $r$ during a revolution period exceeds $\Delta r$ averaged over the period. For fast-changing fields one cannot ignore the deviation of the orbit's shape from a perfect circle.

Our argument relies on the existence of the adiabatic invariant of motion, and in that sense we present another example of an electromagnetic problem that can be treated based on adiabatic invariance (cf. Ref~\onlinecite{crawford_ajp1990}, Sec. XIII-XVII). The argument works for any field magnitude, small or large.  Because of that, we are able to study classical diamagnetism for arbitrary magnitudes of applied fields. In large fields the radius of the circular trajectory changes significantly but its value $r(B)$ is found to be still independent of the history of field variation. After finding the $r(B)$ dependence, we further calculate the induced magnetic moment $\Delta m(B)$. Our solution gives a more satisfying perspective on the textbook limit $B \to 0$, and removes the mysteries associated with the proofs given in Refs.~\onlinecite{textbook_purcell,odell_ajp1986,boyer_ajp2019}.

\section{Diamagnetic atom in arbitrary magnetic field}

Electron's motion is governed by the Coulomb force of the nucleus (charge $Q > 0$, located at the origin), the Lorentz force, and the force applied by the induced electric field. It is, however, useful to perform calculations assuming an arbitrary central potential $V(r)$. We will revert back to the Coulomb potential $V_C = - eQ/r$ later in the section. Newton's equation of motion reads
\begin{equation}\label{eq:Newtons_equation}
m \vec a =
-\frac{\partial V}{\partial r} \hat r - \frac{e}{c} [{\vec v} \times {\vec B(t)}] - e {\vec E}_{ind}(r,t) \ ,
\end{equation}
where $\hat r$ is the unit vector along the electron's radius-vector.

\subsection{Electric field induced by time-dependent magnetic field}

The induced electric field ${\vec E}_{ind}$ is created by the time variation of magnetic field and has to be found from the Maxwell's equations. In general, a spatially-uniform but time-varying field ${\vec B} = B(t)\hat z$ produces not only the electric field ${\vec E}_{ind}(r,t)$ but also some additional, spatially non-uniform ${\vec B}_{ind} = B_{ind}(r,t)\hat z$ \cite{heims_rmp1962, batell_ajp2003}. In a particularly simple special case of linear time dependence $\dot B(t) =$ const, ${\vec B}_{ind}$ exactly vanishes. In that case, assuming that  ${\vec B}$ itself is produced by a solenoid of very large radius with its axis going through the coordinate origin, the exact solution is
\begin{equation}\label{eq:induced_E}
{\vec E}_{ind}(r,t) = - \frac{r \dot B(t)}{2 c} \hat\theta \ ,
\end{equation}
where $\hat\theta$ is the unit vector of polar coordinates $(r,\theta)$ in the $(x,y)$ plane. For a general case, it is further shown in Ref.~\onlinecite{heims_rmp1962} that the induced electric field can be presented as an expansion in powers of $1/c$
$$
{\vec E}_{ind}(r,t) = - \left( \frac{\dot B(t) r}{2 c} + \frac{\dddot B(t) r^3}{16 c^3} + \ldots \right) \hat\theta
$$
Whether corrections to expression (\ref{eq:induced_E}) are significant depends on the magnitudes of time derivatives, and on the observation point. In our case the latter is given by the radius $r_0$ of the atom. For slowly varying function $B(t)$ we can further assume that all characteristic frequencies of the magnetic field time variation are below some $\omega_B$, so that $d^{n}B/dt^n \sim \omega_B^n$. Then we obtain an expansion in a dimensionless parameter $\omega_B^2 r_0^2/c^2$, which can be evaluated for an actual atom. Using the ground state parameters of Hydrogen, where $r_0$ equals Bohr radius $r_B \approx 5 \cdot 10^{-11}$ m, one finds
$$
\frac{\omega_B^2 r_0^2}{c^2} = \left( \frac{\omega_B}{\omega_{B0}} \right)^2 \ ,
\quad
\omega_{B0}  \sim  10^{18} \ {\rm 1/s} \ .
$$
That shows that for an atomic-size object formula (\ref{eq:induced_E}) is applicable with enormous accuracy for all realistic experiments. We will use it in all calculations below.

Nevertheless, for the hypothetical studies of magnetic response in systems with larger $r_0$, it is useful to remember that generally the adiabatic nature of $B(t)$ plays a dual role in the classical diamagnetism problem: the field variation should be slow enough to enable approximate treatments of both the mechanical and the electromagnetic equations. ``Mechanical adiabaticity'' and ``electromagnetic adiabaticity'' are generally two different conditions.

\subsection{Reduction to one-dimensional problem}

Fig.~\ref{fig:atom} places the center of electron orbit at the origin. For that position of the atom, re-writing Eq.~(\ref{eq:Newtons_equation}) in polar coordinates, one gets a system of equations describing electron motion
\begin{eqnarray}
\label{eq:polar_eq_r}
m (\ddot r - r \dot\theta^2) &=& -\frac{\partial V}{\partial r} - \frac{e B}{c} r \dot\theta \ ,
\\
\nonumber
m (r \ddot\theta + 2 \dot r \dot\theta) &=& \frac{e}{c} \left( \frac{r \dot B}{2} + \dot r B \right) \ .
\end{eqnarray}

One may worry that we have significantly restricted our considerations to a single special position of the atomic nucleus on the solenoid axis, and that perhaps other atoms, located away from the axis, may demonstrate a different behavior. However, that is not the case. Using formula (\ref{eq:induced_E}), one can show that away form the axis electric field can be represented as a sum of a spatially uniform component  and a circular component, swirling around the nucleus. The latter is given by the same formula (\ref{eq:induced_E}) but in the polar coordinate system centered at the nucleus. By themselves, the circular and the uniform components produce the diamagnetic and Stark effects. When two components act together, each generally contributes to both effects \cite{ciftja_physlett2024}. However, for atomic radii of the orbits and atomic angular velocities, the influence of the constant field on magnetic response turns out to be negligible. Physically that happens because the constant component alternates along the orbit, while the circular component has the same sign. Since circular components are the same for both atoms, the diamagnetic responses will be approximately equal.

Defining the instantaneous Larmor angular frequency $\Omega_L(t) = e B(t)/ 2mc$, and introducing dimensionless distance $\rho = r/r_0$, one casts the equations of motion in the form
\begin{eqnarray}
\label{eq:polar_eq_rho}
\ddot \rho - \rho \dot\theta^2 &=& -\frac{1}{m r_0^2}\frac{\partial V}{\partial\rho} - 2 \Omega_L \rho \dot\theta \ ,
\\
\label{eq:polar_eq_theta}
\rho \ddot\theta + 2 \dot \rho \dot\theta &=& {\dot\Omega}_L \rho + 2 \Omega_L \dot\rho \ .
\end{eqnarray}
Equation~(\ref{eq:polar_eq_theta})  is equivalent to
\begin{equation}
\label{eq:first_integral}
\frac{d}{dt} \left[ \rho^2 \left(\dot\theta - \Omega_L \right) \right] = 0 \ ,
\end{equation}
implying the existence of the conserved quantity $\lambda = \rho^2 (\dot\theta - \Omega_L(t)) = $ const. Note that conservation of $\lambda$ is based on formula (\ref{eq:induced_E}) and thus assumes electromagnetic adiabaticity. However, it does not require mechanical adiabaticity.

For our purposes, property (\ref{eq:first_integral}) is simply a mathematical fact. However, one could have systematically predicted the existence of that conserved quantity by considering the Lagrangian description of electron motion. If magnetic field is described by the symmetric gauge vector potential $\vec A = (1/2)B(t)r \hat\theta$, the Lagrangian $L(r, \dot r, \theta, \dot\theta)$ turns out to be independent of $\theta$, thus implying a conserved ``canonical angular momentum'' $\Lambda = \partial L/\partial\dot\theta$. The latter turns out to differ only by a factor from the conserved quantity found here, $\Lambda = m r_0^2 \lambda$. The issue of angular momentum conservation in magnetic field \cite{stamps_prl2014}, and the role of gauge choice \cite{textbook_weinberg} is an interesting separate subject that we just touched in our discussion.

We now express
\begin{equation}\label{eq:dot_theta_instantaneous}
\dot\theta = \frac{\lambda}{\rho^2} + \Omega_L(t) \ ,
\end{equation}
and using that in Eq.~(\ref{eq:polar_eq_rho}) obtain an equation
$$
\ddot \rho =  -\frac{1}{m r_0^2}\frac{\partial V}{\partial\rho} + \frac{\lambda^2}{\rho^3} - \rho \Omega_L^2 \ ,
$$
which can be cast in the form of a one-dimensional (1D) Newton equation for a particle of unit mass moving in a time-dependent effective potential
\begin{eqnarray}
\label{eq:effective_radial_equation_dimensionless}
\ddot \rho &=& - \frac{\partial V_{\rm eff}(\rho,t)}{\partial \rho} \ ,
\\
\label{eq:effective_potential_dimensionless}
V_{\rm eff} &=& \frac{V(\rho)}{m r_0^2} + \frac{\lambda^2}{2 \rho^2} + \frac{\Omega^2_L(t) \rho^2}{2}  \ .
\end{eqnarray}
The value of $\lambda$ is defined by the characteristics of the electron orbit in the absence of magnetic field, i.e., at $\Omega_L = 0$. The assumed circular nature of that orbit corresponds to having $\rho = 1$, $\dot\rho = 0$. Eqs.~(\ref{eq:effective_radial_equation_dimensionless}, \ref{eq:effective_potential_dimensionless}) then give
\begin{equation}\label{eq:initial_orbit}
\frac{\partial V_{\rm eff}(\rho)}{\partial \rho} = 0 \quad \Rightarrow \quad
\lambda^2 = \frac{1}{m r_0^2} \left.\frac{\partial V}{\partial\rho}\right|_{\rho = 1}
\end{equation}
Additionally, Eq.~(\ref{eq:dot_theta_instantaneous}) gives $\dot\theta = \lambda$ on that same circular orbit, i.e., $\lambda$ is the angular velocity $\omega_0$ of the electron before the magnetic field application, and will be so denoted from now on.

Before the application of magnetic field, the effective particle stays at rest at the extremum point of the potential $V_{\rm eff}|_{B=0}$. We will see below that in the case of the Coulomb potential $V_{\rm eff}$ has only one extremum, which is a minimum. Thus the effective particle initially resides in a state of stable mechanical equilibrium at the bottom of a potential well. In the following we will be analyzing only such cases. Our analysis will not apply to the maxima or inflection points of $V_{\rm eff}$.

As magnetic field is applied, the shape of $V_{\rm eff}$ changes, setting the effective particle in motion. In the next section the mechanical analogy implied by Eq.~(\ref{eq:effective_radial_equation_dimensionless}) will help us understand the important qualitative features of that motion.

\subsection{Adiabatic variation of magnetic field} \label{sec:adiabatic_invariance}

\begin{figure}[b]
\center
\includegraphics[width = 0.42\textwidth]{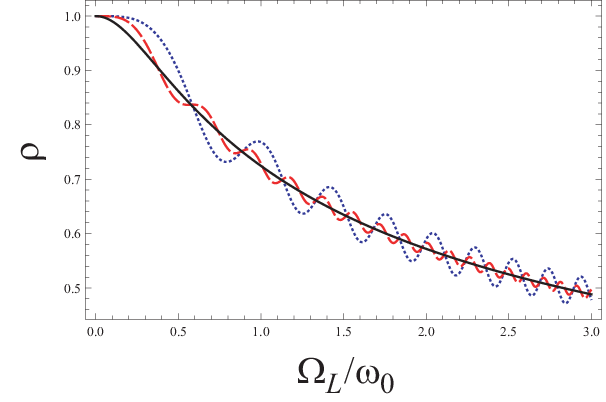}
\caption{Solutions $\rho(t)$ plotted vs.~$\Omega_L(t)/\omega_0$ for the case of Coulomb potential (see Sec.~\ref{seq:Coulomb_potential}) with magnetic field increasing linearly as $\Omega_L(t)/\omega_0 = \alpha \omega_0 t$. Dotted blue line: numeric simulation for $\alpha = 0.2$; Dashed red line: for $\alpha = 0.1$; Solid black line: adiabatic approximation $\rho_{eq}(\Omega_L)$.}
\label{fig:exact_vs_adiabatic_approximation}
\end{figure}

For time-dependent $\Omega_L(t)$ the shape of the effective potential, and the location of its minimum $\rho_{eq}(\Omega_L)$, change with time. In terms of the mechanical analogy it is intuitively clear that when effective potential changes slowly the effective particle will attempt to follow the equilibrium position at the potential's minimum. It is also clear that the particle will not be able to follow $\rho_{eq}$ exactly. In order to chase the moving equilibrium it needs to lag behind, so as to create a force pushing it to the new equilibrium position. The trajectory can be described as $\rho(t) = \rho_{eq}(\Omega_L(t)) + \xi(t)$, and our mechanical intuition suggests that the slower is the change of $\Omega_L(t)$, the smaller will be the deviation $\xi$. Figure~\ref{fig:exact_vs_adiabatic_approximation} illustrates that via a numeric calculation. It compares two simulated solutions $\rho(t)$ for linearly increasing magnetic fields with the approximation $\rho_{eq}(\Omega_L(t))$. Observed deviations are indeed small and progressively decreasing for the slower rates of field change.

A closer look at the situation, however, poses an important question. The intuitive picture of a particle following a potential minimum may strongly rely on the implicit assumption of non-zero friction  in the system. After all, it is the ubiquitous presence of friction that forces particles to come to rest at the bottoms of potential wells. Without friction, perpetual oscillations would be a norm of life, and our intuition would have been quite different. Since there is no friction in Eq.~(\ref{eq:effective_radial_equation_dimensionless}), the ``intuitive'' argument is not that ironclad.

Let's approach this issue by first discussing why one may suspect that a frictionless particle could move away from the energy minimum even if the changes of $V_{eff}$ are small. For that, approximate the equation of motion (\ref{eq:effective_radial_equation_dimensionless}) by Taylor expanding around the moving equilibrium position $\rho_{eq}(t)$. This gives
$$
{\ddot\rho}_{eq} + \ddot{\xi} = - \frac{\partial V_{\rm eff}\big(\rho_{eq} + \xi \big)}{\partial\rho}
\approx - \frac{\partial^2 V_{\rm eff}\big(\rho_{eq} \big)}{\partial\rho^2} \xi \ ,
$$
where we used $\partial V_{\rm eff}/\partial\rho = 0$ at $\rho_{eq}$. Denoting the second derivative of $V_{\rm eff}$ at equilibrium as $\omega^2_R$, we get
\begin{equation} \label{eq:approximate_equation_for_oscillations}
\ddot{\xi} = - \omega^2_R(t) \xi + f(t) \ ,
\end{equation}
i.e., the oscillator equation with variable resonance frequency under the action of applied force
\begin{equation} \label{eq:exciting_force}
f(t) \equiv -\ddot\rho_{eq} = -\frac{d^2 \rho_{eq}}{d\Omega_L^2} {\dot\Omega}_L^2 -  \frac{d \rho_{eq}}{d\Omega_L} {\ddot\Omega}_L \ .
\end{equation}

Adiabatic nature of the magnetic field time dependence is mathematically introduced by writing the  Larmor frequency as $\Omega_L = \Omega_L(\epsilon t)$ where $\Omega_L(x)$ is a smooth function and $\epsilon \to 0$ is a parameter controlling the slowness of the magnetic field variation. This way $V_{eff} = V_{eff}(\rho,\epsilon t)$ and thus the frequency in the linearized equation (\ref{eq:approximate_equation_for_oscillations}) is also a slowly changing function of time $\omega_R = \omega_R(\epsilon t)$. Every time a derivative $d\Omega_L/dt$ is calculated, one gets a factor of $\epsilon$. According to Eq.~(\ref{eq:exciting_force}), this means the force $f(t)$ is small.

Equation (\ref{eq:approximate_equation_for_oscillations}) can be discussed qualitatively from the physics point of view. The small force $f(t)$ gently agitates the particle. However, the smallness of the force is not enough to ensure the smallness of the response $\xi$. In the absence of friction a small but resonant force may produce a large deviation from equilibrium. This scenario can be avoided if we require $f(t)$ to have characteristic frequencies well below the current resonance frequency $\omega_R(t)$. One hopes to be able to achieve that if $f(t)$ can be made to vary as slow as needed.

The above conditions, imposed on $f(t)$, are still not sufficient for obtaining a small $\xi$. Yet another scenario capable of producing a large $\xi$ is realized when the restoring force coefficient $\omega_R^2(t)$ varies---even with small amplitude---in sync with the particle's oscillations. That is the case of ``parametric amplification'' (see Ref.~\onlinecite{textbook_hand_finch}, Ch. 10, Sec. 10.2). Again, it is logical to assume that parametric resonance can be avoided if $\omega_R(t)$ is made to vary as slow as needed.

Qualitative expectations outlined above find a beautiful confirmation in a very general result of classical mechanics---the existence of an adiabatic invariant for the motion of a particle in a potential that depends on a slowly changing parameter \cite{crawford_ajp1990, textbook_landau_lifshitz, textbook_goldstein}, \cite{textbook_hand_finch}(Ch.~6, Sec. 6.5). Equations~(\ref{eq:effective_radial_equation_dimensionless}, \ref{eq:effective_potential_dimensionless}) constitute a particular example of a situation where adiabatic invariance is observed.

The theory of adiabatic invariants is an advanced subject but in the present one-dimensional case its predictions can be formulated quite simply. The integral
$$
J = \frac{1}{2\pi} \oint p dq \ ,
$$
where $p$ is the particle's momentum and $q$ is its 1D coordinate, taken in the configuration space $(q,p)$ over the closed path representing particle's oscillations around the equilibrium at a fixed parameter value, is almost conserved \cite{textbook_landau_lifshitz, textbook_goldstein, textbook_hand_finch}. The slower is the time change of the parameter, the smaller will be the deviation of $J$ from its initial value. Qualitatively, the change of parameters can be considered slow if the frequency of oscillations changes very little during one oscillation period. But the precise formulation is beyond the scope of the present discussion \cite{textbook_landau_lifshitz}.

\begin{figure}[t]
\center
\includegraphics[width = 0.35\textwidth]{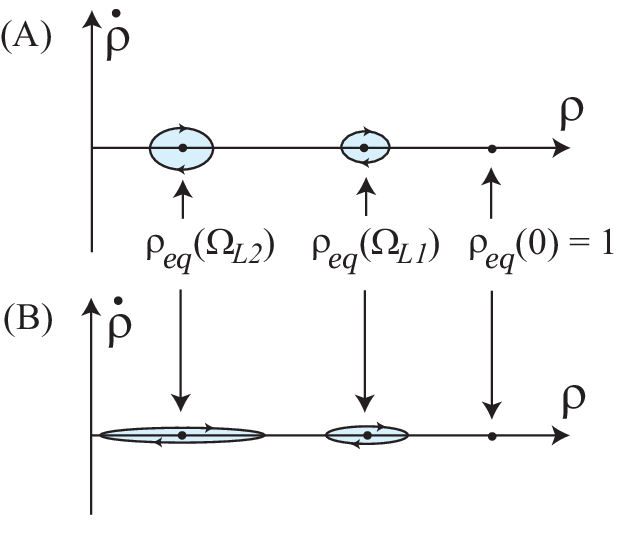}
\caption{(A) Path evolution in phase space: the areas enclosed by the paths change but remain small; (B) Another path evolution where enclosed areas remain small.}
\label{fig:phase_portraits_sketch}
\end{figure}

Specializing the general 1D expression above to the case of Eqs.~(\ref{eq:effective_radial_equation_dimensionless}, \ref{eq:effective_potential_dimensionless}) with unit mass of effective particle, one gets $q = \rho$, $p = \dot\rho$, and
\begin{equation}\label{eq:J_definition}
J = \frac{1}{2\pi} \oint \dot\rho d\rho \ ,
\end{equation}
where integration is performed in the $(\rho,\dot\rho)$ phase space. In Eq.~(\ref{eq:J_definition}) the integral on the right equals the area inside the closed path. Fig.~\ref{fig:phase_portraits_sketch}(A) shows a sketch of the evolution of the closed path after the magnetic field is turned on. At $\Omega_L = 0$ the effective particle sits exactly at the potential minimum, the amplitude of oscillations is zero, and the path degenerates to a single point, giving $J|_{\Omega_L = 0} = 0$. As the field is slowly varied, the area encircled by the path may slightly change but must remain close to its original zero value, as $J$ is an adiabatic invariant. Note that the whole argument applies to the full potential $V_{eff}$, without the need of linearization that produced Eq.~(\ref{eq:approximate_equation_for_oscillations}). The equilibrium point $\rho_{eq}(\Omega_L)$ moves with changing magnetic field. The closed path follows and always encircles it.

To go from the conservation of $J$ to the desired statement about effective particle staying near the moving potential energy minimum, we need to overcome one additional objection. Consider the sequence of path transformations illustrated in Fig.~\ref{fig:phase_portraits_sketch}(B). The shape of the path changes in such a way that while the enclosed area is still conserved, the range of motion along the $\rho$-axis increases, potentially becoming comparable with $\rho_{eq}$ itself. If such a scenario were possible, the conservation of $J$ would not be sufficient to prove that the particle closely follows the minimum. However, one can show that such situation is, in fact, ruled out. The path in the definition of $J$ represents the motion of the particle at a fixed parameter value. In such motion the total energy $E = \dot\rho^2/2 + V_{eff}(\rho)$ is conserved. Near the equilibrium that gives
\begin{equation}\label{eq:energy_conservation_near_equilibrium}
\frac{\dot\rho^2}{2} + \frac{\omega_R^2 (\rho - \rho_{eq})^2}{2} =  E = {\rm const},
\end{equation}
and consequently the maximum deviation from equilibrium and the maximum velocity satisfy
$$
\omega_R {|\rho - \rho_{eq}|_{*}} = |\dot\rho|_{*} = \sqrt{2E} \ .
$$
Equation (\ref{eq:energy_conservation_near_equilibrium}) additionally means that in the limit of small oscillations the shape of the path is an ellipse, so
$$
J = \frac{1}{2\pi} \big( \pi |\rho - \rho_{eq}|_{*} \cdot |\dot\rho|_{*} \big) =
\frac{\omega_R (|\rho - \rho_{eq}|_{*})^2}{2},
$$
and
\begin{equation}\label{eq:max_deviation}
|\rho - \rho_{eq}|_{*} = \sqrt{\frac{2 J}{\omega_R}} \ .
\end{equation}
The last equation shows that the situation depicted in Fig.~\ref{fig:phase_portraits_sketch}(B) cannot happen as long as $\omega_R$ does not drop to zero. More formally, for any finite $\omega_R$ a sufficiently slow change of $\Omega_L(t)$ will make the deviation $|\rho - \rho_{eq}|_{*}$ arbitrarily small.

It should be mentioned that conservation of $J$ is only valid when effective particle oscillates around the minimum of $V_{eff}$. Without giving a proof, we warn that according to the general theory the maxima and saddle points of potential cannot be treated that way  \cite{textbook_landau_lifshitz, textbook_goldstein, textbook_hand_finch}.

As we see, adiabatic invariance is a very powerful statement that proves the desired property of solution without the need to compute the solution itself. An alternative and maybe more explicit way to reach the same conclusion is discussed in Appendix~\ref{Appendix:proof_of_small_deviations}, where the solution of the linearized equation (\ref{eq:approximate_equation_for_oscillations}) is obtained.

\subsection{Case of Coulomb potential} \label{seq:Coulomb_potential} 

\begin{figure}[b]
\center
\includegraphics[width = 0.47\textwidth]{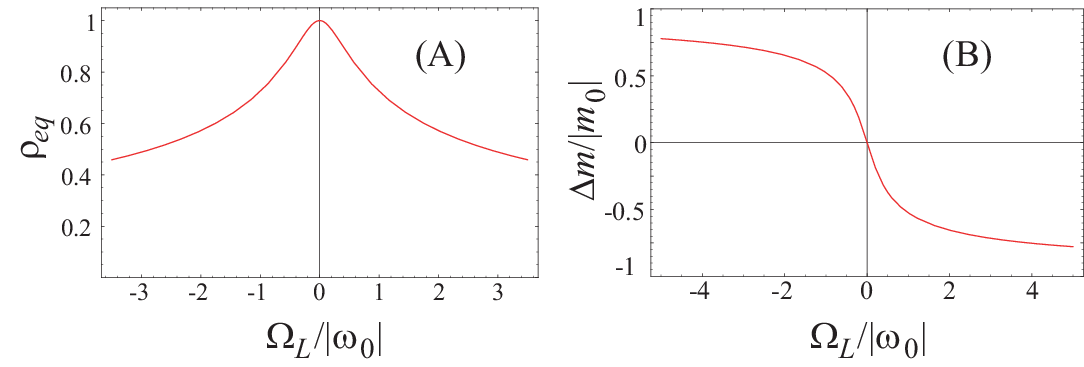}
\caption{(A) Position $\rho_{eq}$ of the minimum of effective potential $V^C_{\rm eff}$, as a function of the dimensionless frequency $\Omega_L/|\omega_0|$. (B)~Normalized field-induced magnetic moment change $\Delta m/|m_0|$ as a function of $\Omega_L/|\omega_0|$.}
\label{fig:reqm}
\end{figure}

For the Coulomb potential one has $V_C/m r_0^2 = - e Q/ m \rho r_0^3$ and Eq.~(\ref{eq:initial_orbit}) gives $e Q/m r_0^3 = \omega_0^2$. From there, using notation $W_L = \Omega_L/\omega_0$, one gets
\begin{equation}\label{eq:effective_VCoulomb}
V^C_{\rm eff} = \omega_0^2 \left(-\frac{1}{\rho} + \frac{1}{2\rho^2} + \frac{W_L^2 \rho^2}{2} \right) \ .
\end{equation}
Analyzing this formula one can see that for $\rho > 0$ effective potential $V^C_{\rm eff}$ indeed has only one stationary point that is a minimum. Its position $\rho_{eq}(W_L)$ is determined by the equation
\begin{equation}\label{eq:minimum_position}
\rho_{eq} + W_L^2 \rho_{eq}^4 = 1 \ .
\end{equation}
Since $W_L$ enters this equation through its square, $\rho_{eq}(W_L)$ will be an even function. Figure~\ref{fig:reqm}(A) shows its graph, obtained numerically. Equilibrium radius decreases with growing $|W_L|$, regardless of the sign of the applied field. Analytic approximations can be found in the limits of small and large values of $W_L$
\begin{eqnarray}
\label{eq:limit_rho_small_w}
\rho_{eq} & \approx & 1 - W_L^2 \qquad \qquad \quad (|W_L| \ll 1) \ ,
\\
\label{eq:limit_rho_large_w}
\rho_{eq} & \approx & \frac{1}{\sqrt{|W_L|}} - \frac{1}{4|W_L|}  \quad (|W_L| \gg 1) \ .
\end{eqnarray}
Formula (\ref{eq:limit_rho_small_w}) justifies the assumption of constancy of radius at small applied fields that was discussed in the Introduction.

Frequency $\omega_R(\rho_{eq})$ is found as
$$
\omega^2_R = \left. \frac{\partial^2 V_{\rm eff}^C}{\partial \rho^2} \right|_{\rho_{eq}} = \omega_0^2 \left( \frac{3}{\rho_{eq}^4} - \frac{2}{\rho_{eq}^3}  + W_L^2 \right) \ .
$$
Using the relation (\ref{eq:minimum_position}) we express $W_L^2 = (1-\rho_{eq})/\rho_{eq}^4$ and obtain
\begin{equation}\label{eq:omega_R}
\omega^2_R = \omega_0^2 \ \frac{4 - 3 \rho_{eq}}{\rho_{eq}^4} \ .
\end{equation}
In the range $\rho_{eq} \in (0, 1)$ that frequency is monotonically decreasing from infinity at $\rho_{eq} = 0$ to $\omega_R = \omega_0$ at $\rho_{eq} = 1$, so $\omega_R \geq |\omega_0|$ holds in all fields. The requirement of non-zero $\omega_R$ is thus satisfied, and the analysis of the previous section fully applies.  Expression (\ref{eq:max_deviation}) for the maximum deviation from the potential minimum specializes to $|\rho - \rho_{eq}|_{*} \leq \sqrt{2 J/\omega_0}$.

Qualitative condition of mechanical adiabaticity discussed in Sec.~\ref{sec:adiabatic_invariance} is expressed as
$$
\left| \frac{d \omega_R}{d W_L} \right| \cdot \left| \frac{d W_L}{dt} \right| T \ll \omega_R
$$
with period $T = 2\pi/\omega_R$. In other words,
\begin{equation}\label{eq:mechanical_adiabaticity_condition_general}
\left| \frac{d W_L}{dt} \right|  \ll \frac{1}{2\pi} \frac{\omega_R^2}{\left| d \omega_R/d W_L \right| }
\end{equation}
has to be satisfied. The right hand side of that inequality can be expressed as a function of $\rho_{eq}$ using equations (\ref{eq:minimum_position}) and (\ref{eq:omega_R}). This calculation gives
$$
\left| \frac{d W_L}{dt}  \right| \ll \frac{|\omega_0|}{2\pi}
\left( \frac{(4-3\rho_{eq})^{5/2}}{\rho_{eq}^4 (16 - 9 \rho_{eq}) \sqrt{1 - \rho_{eq}}}
\right) .
$$
The expression in parenthesis on the right diverges for $\rho_{eq} \to 0$ (i.e., $|W_L| \to \infty$) and $\rho_{eq} \to 1$ (i.e., $W_L \to 0$) and has a single minimum in the interval $(0,1)$ located at $\rho_{eq} \approx 0.945$, where its value is approximately 1.045 and magnetic field corresponds to $|W_L| \approx 0.26$. As a result, to maintain the mechanical adiabaticity condition for the whole field range from zero to infinity one needs $dW_L/dt \ll \omega_0/2\pi$, or equivalently
\begin{equation}\label{eq:mechanical_adiabaticity_condition_Coulomb}
\left| \frac{d \Omega_L}{dt} \right| \ll \frac{\omega_0^2}{2\pi} \  \ .
\end{equation}
The analysis above is consistent with the numeric results shown in Fig.~\ref{fig:exact_vs_adiabatic_approximation}. For the linear dependence $W_L = \alpha \omega_0 t$ used in that example, condition (\ref{eq:mechanical_adiabaticity_condition_Coulomb}) translates to $\alpha \ll 1/2\pi \approx 0.16$. For the choices of $\alpha = 0.1$ and 0.2 in Fig.~\ref{fig:exact_vs_adiabatic_approximation}, the strong inequality is not satisfied and we indeed see noticeable deviations from the adiabatic approximation.

Angular velocity $\omega_0$ can be evaluated for the electron in the ground state orbit of a Hydrogen atom. With $r_0 = r_B \approx 5 \cdot 10^{-11}$ m one gets $\omega_0 \approx 4.1 \cdot 10^{16} \ {\rm 1/s}$. Just like the condition of electromagnetic adiabaticity,  mechanical adiabaticity condition is satisfied in almost all realistic experiments with atomic-size systems. For example, if one could ramp magnetic field from zero to 10 Tesla ($\Omega_L = 8.8 \cdot 10^{11}$ 1/s) in 1 nanosecond, condition (\ref{eq:mechanical_adiabaticity_condition_Coulomb}) would still be well satisfied.

\subsection{Magnetic moment}  
\label{sec:magentic_moment}

For an electron moving with constant angular velocity $\omega$ along a circular orbit of radius $r$, magnetic moment is given by $m = (- e) r^2 \omega/2$. In the case of a slowly changing orbit, the moment must be defined by averaging over the rotation period
$$
m(t) = \frac{(-e) \langle r^2(t) \, \dot\theta(t) \rangle }{2} \ .
$$
In the adiabatic approximation one can use $r = r_0 \rho_{eq}(W_L)$ for the radius, and the slowly varying $\dot\theta$ is given by formula ({\ref{eq:dot_theta_instantaneous}). Altogether this gives
\begin{eqnarray}
\nonumber
m(W_L) &=&  m_0 \big( 1 + W_L \, \rho_{eq}^2(W_L) \big) \ ,
\\
\nonumber
m_0 &=& \frac{(-e) r_0^2 \omega_0 }{2} \ .
\end{eqnarray}
Note that for $\omega_0 > 0$, as in Fig.~\ref{fig:atom}, the initial magnetic moment is negative, which also affects the sign of $W_L$. To reduce confusion with signs, it is convenient to express the field-induced change of magnetic moment in terms of $\Omega_L$, which by definition has the same sign as $B$. One finds
\begin{equation}\label{eq:Delta_m}
\Delta m = -\left( \frac{er_0^2}{2} \right) \Omega_L \rho_{eq}^2 (\Omega_L) \ .
\end{equation}
Magnetic response has the correct diamagnetic sign and is given by an odd function of $\Omega_L$. At large fields, using approximation (\ref{eq:limit_rho_large_w}) in Eq.~(\ref{eq:Delta_m}) we find  $\Delta m \to \mp |m_0|$ for $\Omega_L \to \pm\infty$, i.e., the initial moment can be either fully suppressed, or doubled. At small fields, using approximation (\ref{eq:limit_rho_small_w}) we find $\Delta m \approx m_0 W_L = -(e^2 r_0^2/4 c m)B$, in accord with the standard expression (\ref{eq:m_small_B}). The full graph of $\Delta m(\Omega_L)$ is shown in Fig.~\ref{fig:reqm}B.

\subsection{Electron in harmonic potential}
It is instructive to consider electron moving in a quadratic (harmonic) central potential $V = k r^2/2$ instead of the Coulomb potential. In that case the effective potential (\ref{eq:effective_potential_dimensionless}) is given by
$$
V_{\rm eff} = \frac{\Omega_0^2 \rho^2}{2} + \frac{\omega_0^2}{2\rho^2} + \frac{\Omega_L^2 \rho^2}{2} \ ,
$$
where $\Omega_0^2 = k/m$. Potential's minimum is located at
$$
\rho_{eq}^4 = \frac{\omega_0^2}{\Omega_0^2 + \Omega_L^2} \ ,
$$
and small oscillations around it have the frequency
$$
\omega_R^2 = \Omega_0^2 + \Omega_L^2 + \frac{3 \omega_0^2}{\rho_{eq}^4}
= 4 (\Omega_0^2 + \Omega_L^2) \ .
$$
Since $\omega_R^2 \geq 4 \Omega_0^2$ is always positive, the adiabatic invariant approach is applicable.

Angular velocity of electron rotation (\ref{eq:dot_theta_instantaneous}) assumes the value
$$
\frac{\dot\theta}{\omega_0} = \frac{1}{\rho_{eq}^2} + \frac{\Omega_L}{\omega_0}
= \sqrt{\frac{\Omega_0^2 + \Omega_L^2}{\omega_0^2}} + \frac{\Omega_L}{\omega_0} \ .
$$
The field-induced change of magnetic moment is then given by
$$
\Delta m = - m_0 \frac{\Omega_L}{\sqrt{\Omega_0^2 + \Omega_L^2}}  \ .
$$
The graph of magnetization change looks qualitatively similar to that in Fig.~\ref{fig:reqm}(B). However, the transition from the linear in field behavior to the saturation behavior now happens at $\Omega_L \sim \Omega_0$, instead of the $\Omega_L \sim \omega_0$ in the Coulomb case.

\section{Discussion}  
Our main goal was to provide a theoretical justification for the persistence of the circular nature of atomic electron's orbit as the external magnetic field is turned on. It was shown that persistence of circularity follows from the two key facts. First, the conserved quantity (\ref{eq:first_integral}) reduces the problem to a one-dimensional equation of motion (\ref{eq:effective_radial_equation_dimensionless}). Second, the adiabatic invariant (\ref{eq:J_definition}) explains why the effective particle remains near the minimum of a slowly varying potential, even in the absence of friction. Once the circularity of the orbit is established, the solution for an arbitrary field strength becomes straightforward and transparent.

Two observations can be made based on our solution. First, the radius of the orbit $r_0 \rho_{eq}(\Omega_L)$ turned out to be an even function of the applied field. This is remarkable because there is an obvious asymmetry between the electron's motions for positive and negative magnetic fields. For one sign of $B$, the induced electric field $\vec E_{ind}$ accelerates the electron, while for the other it attempts to stop it~\cite{footnote:electron_does_not_stop}. Accordingly, the electron's angular velocity dependence on $B$ is asymmetric, with asymmetry being especially pronounced in the non-linear regime (Fig.~\ref{fig:angular_velocity_vs_w}). Nevertheless, $\rho_{eq}(\Omega_L)$ turns out to be an even function, and this property holds not just for the Coulomb but for any radial potential $V(r)$, since according to in Eq.~(\ref{eq:effective_potential_dimensionless}) $V_{\rm eff}$ is always a function of $\Omega_L^2$. The even nature of $\rho_{eq}(\Omega_L)$ in turn guarantees the odd nature of $\Delta m(\Omega_L)$: calculations of Sec.~\ref{sec:magentic_moment} rely only on Eq.~({\ref{eq:dot_theta_instantaneous}) that does not depend on the form of $V(r)$.

\begin{figure}[t]
\center
\includegraphics[width = 0.47\textwidth]{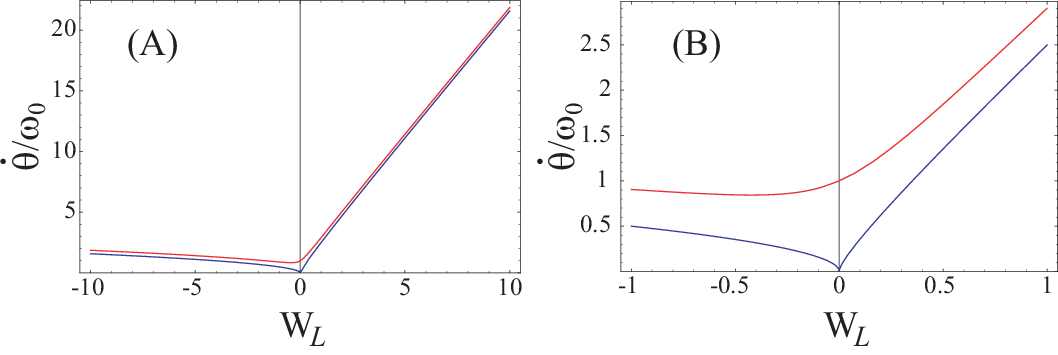}
\caption{Angular velocity of the orbiting electron $\dot\theta(W_L)/\omega_0$ in applied field. (A) wide-range view and (B) region near zero. Red: adiabatic approximation. Blue: approximation $W_L + |W_L| + \frac{1}{2}\sqrt{|W_L|}$ that follows from expression (\ref{eq:limit_rho_large_w}).}
\label{fig:angular_velocity_vs_w}
\end{figure}

Second, an interesting perspective on our solution can be obtained from the point of view of Larmor's theorem. In its basic version, the theorem states that the trajectory of electron in a constant magnetic field $r_B(t)$ can be obtained by uniformly rotating its zero-field trajectory $r(t)$ with angular velocity $\Omega_L$ \cite{feynman_lectures}. A generalization of the Larmor's theorem to time dependent fields \cite{heims_rmp1962} states that if the field is continuously increased from zero to $B$, electron's trajectory will continuously transform from $r(t)$ to $r_B(t)$. In the latter form the theorem is directly applicable to electron's motion in a central potential and states that with changing magnetic field the angular velocity changes as $\omega_0 \to \omega_0 + \Omega_L$. Importantly, the Larmor's theorem (in either form) is only applicable in weak fields, i.e., when the Lorentz force is much smaller than the force due to the central potential, or $\Omega_L \ll \omega_0$ \cite{feynman_lectures, textbook_zangwill}. We can now check that our adiabatic approximation solution is consistent with the theorem's statement. Using Eq.~(\ref{eq:dot_theta_instantaneous}) we can write
\begin{equation}\label{eq:Larmor_analogue}
\frac{\partial \dot\theta}{\partial \Omega_L} =
- \frac{2 \omega_0}{\rho^3} \frac{\partial \rho_{eq}}{\partial\Omega_L} + 1 \ .
\end{equation}
At $\Omega_L = 0$ the derivative $\partial\rho_{eq}/\partial\Omega_L$ vanishes, and we get the equivalent of Larmor's theorem: $d\dot\theta = d\Omega_L$. As $|\Omega_L|$ is increased, the first term in Eq.~(\ref{eq:Larmor_analogue}) acquires non-zero value, and one observes how Larmor's theorem is violated outside of its validity range. In particular, for $\Omega_L \to +\infty$ approximation (\ref{eq:limit_rho_large_w}) gives $\partial\dot\theta/\partial\Omega_L \to 2$. In that limit the angular velocity $\dot\theta$ approaches the cyclotron frequency, $\Omega_c = 2 \Omega_L$, exhibited by a free electron orbiting in external magnetic field. Electron in a diamagnetic atom is not free but for $\Omega_L \to +\infty$ the Coulomb force acting on it becomes negligible in comparison with the Lorentz force.

\section{Practical classroom use}
This paper may be used to guide discussions with senior undergraduates taking an Electricity and Magnetism course at the level of Ref.~\onlinecite{textbook_griffiths}. Minimal prerequisite is a course of Newtonian mechanics and familiarity with the phase space. At that level the adiabatic invariant may be introduced as a useful result of advanced analytical mechanics that can be invoked without proof.

For students with exposure to courses in analytical mechanics, including junior graduate students, the problem of classical diamagnetism may serve as a motivator for the introduction of the Green's function method and the WKB approximation. This can expose such students to both approaches as general mathematical methods rather than exclusively quantum-mechanical tools.

Teaching of two topics in electromagnetism can benefit from our paper. The first is the classical diamagnetism itself. The second in the Larmor's theorem and its domain of applicability.

This study itself arose from the classroom discussion about the constant radius approximation that is usually suggested as a hint for the classical diamagnetic atom homework problem.

\section{Acknowledgements}
We wish to thank the anonymous Referee for the thoughtful comments that resulted in a better version of this paper.

\appendix

\section{}\label{Appendix:proof_of_small_deviations}

With slow $\Omega_L(\epsilon t)$ Eq.~(\ref{eq:approximate_equation_for_oscillations})  assumes the form
\begin{equation}\label{appA:oscillator_with_slow_time}
\ddot{\xi}(t) = -\omega_R^2(\epsilon t) \xi(t) + f(\epsilon t) \ ,
\end{equation}
where $\omega_R$ and $f$ are said to depend on the ``slow time'' $\tau = \epsilon t$. A homogeneous version ($f = 0$) of that equation is treated in Ref.~\onlinecite{textbook_bender_orszag} (Ch. 11, Sec. 11.3, Example 3). It is shown that in the $\epsilon \to 0$ limit it can be transformed to a form amenable to analysis on the basis of the WKB approximation. Furthermore, when a non-homgeneous term $f$ is present, the solution can be obtained using the WKB Green's function (Ref.~\onlinecite{textbook_bender_orszag}, Ch. 10, Sec. 10.3). In our case with initial conditions at $t = 0$ being $\xi = 0$ and $d{\xi}/dt = 0$, solution can be found using the retarded WKB Green's function
\begin{eqnarray}
\nonumber
G(t,t') &=& \frac{\sin\Phi(t,t')}{\sqrt{\omega_R(\epsilon t) \omega_R(\epsilon t')}}  \quad (t > t') \ ,
\\
\label{appA:WKB_Greens_function}
G(t,t') &=& 0  \quad (t < t') \ ,
\\
\nonumber
\Phi(t,t') &=& \int_{t'}^t \omega_R(\epsilon u) du
\end{eqnarray}
(note how in the case of constant $\omega_R$ these expressions revert to the exact Green's function of a harmonic oscillator, see Ref.~\onlinecite{textbook_hand_finch}, Ch. 3, Sec. 3.6).

Solution is given by the integral
\begin{equation}\label{appA:convolution_with_G}
\xi(t) = \int_0^t G(t,t') f(\epsilon t') dt' \ .
\end{equation}
In this form the possibility of parametric resonance is already excluded, as long as the WKB formula (\ref{appA:WKB_Greens_function}) remains a valid approximation for the Green's function. One necessary condition for that is for $\omega_R(\epsilon t)$ to remain non-zero \cite{textbook_hand_finch}. This requirement also means that function $|G|$ is bounded from above.

The requirement of smallness of the integral (\ref{appA:convolution_with_G}) further leads to two conditions. First, the integrand has to be small. The bounded nature of $G$ means that it is enough to have small $|f|$ to achieve that. Second, one needs to make sure that the integral of a small function remains small even for a very long integration time $t$. This is where a resonant exciting force could present a danger: external force $f(t')$ oscillating in sync with $\sin\Phi(t,t')$ would produce a non-zero average of the integrand and lead to large values of $\xi$ as $t \to \infty$. However, a slow-changing force $f(\epsilon t')$ would not be dangerous. One can prove that by changing the integration variable to $\tau' = \epsilon t'$ in Eq.~(\ref{appA:convolution_with_G}). The factor $\sin\Phi(t,t')$ will become a rapidly oscillating function of $\tau'$, while $f(\tau')$ will remain almost constant on the scale of oscillation periods. The integral for $\xi$ will approach zero for $\epsilon\to 0$ due to the presence of a fast-oscillating factor.

We will now specialize the discussion to the case of the Coulomb potential. Here one finds
\begin{eqnarray*}
\omega^2_R &=& \omega_0^2 \left( \frac{3}{\rho_{eq}^4} - \frac{2}{\rho_{eq}^3}  + W_L^2 \right) \ ,
\\
f &=& - \big( \rho_{eq}'' {\dot W}_L^2 + \rho_{eq}' {\ddot W}_L \big) \ ,
\end{eqnarray*}
where $\rho'_{eq} \equiv d\rho_{eq}/dW_L$ and $\rho''_{eq} \equiv d^2\rho_{eq}/dW_L^2$.
These quantities are found by differentiating Eq.~(\ref{eq:minimum_position}) with respect to $W_L$
\begin{eqnarray*}
\rho_{eq}' &= & - \frac{2W_L\rho_{eq}^4}{1+4W_L^2\rho_{eq}^3}\, ,
\\
\rho_{eq}'' &=& \frac{2\rho_{eq}^4 \big( 24 W_L^4 \rho_{eq}^6 + 8 W_L^2 \rho_{eq}^3 -1 \big)}{(1+4W_L^2\rho_{eq}^3)^3}\, .
\end{eqnarray*}
Further expressing $W_L$ through $\rho_{eq}$ from Eq.~(\ref{eq:minimum_position}), we get
\begin{eqnarray*}
\rho_{eq}' &=& -\frac{2\rho_{eq}^3\sqrt{1-\rho_{eq}}}{4-3\rho_{eq}} \ ,
\\
\rho_{eq}'' &=& \frac{2\rho_{eq}^5\big(15\rho_{eq}^2-40\rho_{eq}+24\big)}{(4-3\rho_{eq})^3} \ .
\end{eqnarray*}

Obtained expressions have no divergences on the interval $\rho_{eq} \in (0,1)$, and are thus bound. Standard process of investigating their extrema shows that $|\rho'_{eq}|$ has a maximum at $\rho_{eq} = (20 - 2 \sqrt{10})/15$, and $|\rho''_{eq}|$ has a maximum at $\rho_{eq} = 1$. That translates into
\begin{eqnarray*}
|\rho'_{eq}| & \leq & 0.356 \ ,
\\
|\rho''_{eq}| & \leq & 2 \ .
\end{eqnarray*}
The existence of the bounds shows that $|f(\epsilon t)|$ can be made uniformly small on the interval $t \in (0,\infty)$. Condition $\omega_R \geq \omega_0$ was already proved in Sec.~\ref{seq:Coulomb_potential}. This way all conditions required for the smallness of $\xi$ are satisfied.

Overall, the conclusion is that a small and slow-changing external force cannot produce large deviations from equilibrium in a frictionless oscillator with slow-changing stiffness.



\begin{thebibliography}{2}


\bibitem{textbook_griffiths}
D. J. Griffiths, {\em Introduction to Electrodynamics}, 4-th ed. (Pearson, Boston, 2013), Sec. 6.1.3.

\bibitem{textbook_purcell}
E. M. Purcell, {\em Electricity and Magnetism}, Berkeley physics course Vol. 2, 2-nd ed. (McGraw-Hill, Inc., New York, 1985), Sec. 11.4.

\bibitem{feynman_lectures}
R. P. Feynman, R. B. Leighton, and M. Sands, {\em The Feynman lectures on Physics}, Vol. II (Basic Books, 2010) Ch.~34.

\bibitem{vanvleck_book_1932}
J. H. Van Vleck, {\em The theory of electric and magnetic susceptibilities}, (Clarendon, Oxford, 1932), Ch. 3, Sec. 24.

\bibitem{textbook_aharoni}
A. Aharoni, {\em Introduction to the Theory of Ferromagnetism} (Oxford University Press, Oxford, 1996), Ch. 1, Sec. 1.3.

\bibitem{textbook_blundell}
S. Blundell, {\em Magnetism in Condensed Matter} (Oxford University Press, Oxford, 2001), Ch.1, Sec. 1.2.2

\bibitem{odell_ajp1986}
S. L. O’Dell and R. K. P. Zia,
``Classical and semiclassical diamagnetism: A critique of treatment in elementary texts'',
Am. J. Phys. {\bf 54}, 32 (1986).

\bibitem{boyer_ajp2019}
T. H. Boyer,
``Diamagnetic behavior in random classical radiation'',
Am. J. Phys. {\bf 87}, 915 (2019).

\bibitem{crawford_ajp1990}
F. S. Crawford,
``Elementary examples of adiabatic invariance'',
Am. J. Phys. {\bf 58}, 337 (1990).

\bibitem{heims_rmp1962}
S. P. Heims and E. T. Jaynes,
``Theory of Gyromagnetic Effects and Some Related Magnetic Phenomena'',
Rev. Mod. Phys. {\bf 34}, 143 (1962).

\bibitem{ciftja_physlett2024}
O. Ciftja and C. L. Bentley Jr.,
``Impact of induced electric field on the motion of a charged particle in a
uniform time-dependent magnetic field that is linearly increasing with time''
Phys. Lett. A {\bf 526}, 129974 (2024).

\bibitem{stamps_prl2014}
C. R. Greenshields, R. L. Stamps, S. Franke-Arnold, and S. M. Barnett,
``Is the Angular Momentum of an Electron Conserved in a Uniform Magnetic Field?'',
Phys. Rev. Lett. {\bf 113}, 240404 (2014).

\bibitem{textbook_weinberg}
S. Weinberg, {\em Lectures on Quantum Mechanics}, 2-nd ed. (Cambridge University Press, Cambridge, 2015), Ch.~9,~10.

\bibitem{footnote:electron_does_not_stop}
Naively, one could think that when $E_{ind}$ continuously decelerates the electron, the latter should eventually stop. However, the radius of the orbit decreases with time, and in accord with Eq.~(\ref{eq:induced_E}) the electron experiences a progressively smaller force $(-e) E_{ind}$. Electron's linear acceleration is always negative but the change of velocity turns out to be finite.

\bibitem{batell_ajp2003}
B. Batell and A. Ferstl,
``Electrically induced magnetic fields: a consistent approach",
Am. J. Phys. {\bf 71}, 925 (2003).

\bibitem{textbook_landau_lifshitz}
L. D. Landau, E. M. Lifshitz, {\em Course of Theoretical Physics, vol. I, Mechanics}, 3-rd ed. (Elsevier, Oxford, 1976), Sec. 49-51.

\bibitem{textbook_goldstein}
H. Goldstein, C. Poole, J. Safko, {\em Classical Mechanics}, 3-rd ed. (Addison Wesley, San Francisco, 2001), Ch. 10, Sec. 10.6.

\bibitem{textbook_hand_finch}
L. N. Hand and J. D. Finch, {\em Analytical Mechanics}, (Cambridge University Press, Cambridge, 1998).

\bibitem{textbook_zangwill}
Andrew Zangwill, {\em Modern Electrodynamics}, 1-st ed. (Cambridge University Press, Cambridge, 2012),
Ch. 12, Sec. 12.2.3.

\bibitem{textbook_bender_orszag}
C. M. Bender and S. A. Orszag, {\em Advanced Mathematical Methods for Scientists and Engineers I}, 2-nd ed. (Springer, New York, 1999).

\end{thebibliography}
\end{document}